\documentstyle[12pt]{article}
\newcommand \be[1]{e_{\bar{\nu}^{\ast}_{#1}}}
\newcommand \ipro[1]{\langle{#1}\rangle}
\newcommand {\bx}{\bar{x}}
\newcommand {\by}{\bar{y}}
\newcommand \w[1]{W_{#1}}

\newcommand \BX[1]{\mbox{\Large $\Box$}_{l^{({#1})}}}
\newcommand {\Kappa}{\mbox{\Large $\kappa$}}

\newcommand {\beq}{\begin{eqnarray}}
\newcommand {\eeq}{\end{eqnarray}}
\newcommand {\beqs}{\begin{eqnarray*}}
\newcommand {\eeqs}{\end{eqnarray*}}
\newcommand {\del}{\partial}

\newcommand \th[1]{{\Theta}_{#1}}
\newcommand {\bp}{\bf P}
\newcommand {\bc}{\bf C}
\newcommand {\bz}{\bf Z}

\newcommand {\vpi}{\varpi}
\newcommand \vvp[1]{{\varpi}_{#1}}

\newcommand {\Om}{\Omega}

\newcommand \gam[1]{\Gamma \left({#1}\right)}
\newcommand \gamm[2]{\Gamma \left({\frac{#1}{#2}}\right)}
\newcommand \dnu[1]{{\nu}^{\ast}_{#1}}
\newcommand \bdnu[1]{{\bar{\nu}}^{\ast}_{#1}}
\newcommand {\cf}{\cal F}

\newcommand {\amb}{{\bf P}_{d+1} [2,2,2,\cdots ,2,2,1,1](2(d+1))}
\newcommand {\poly}{\Delta}
\newcommand {\dpoly}{{\Delta}^{\ast}}
\newcommand {\mfd}{\mbox{manifold}}
\newcommand {\mfds}{\mbox{manifolds}}

\newcommand {\kae}{\mbox{K{\"a}hler}}

\newcommand {\mm}{\cal M}
\newcommand {\mw}{\cal W}
\newcommand{\pr}{\hspace{\parindent}}

\begin{document}
\setlength{\oddsidemargin}{0cm}
\setlength{\baselineskip}{7mm}

\begin{titlepage}
    \begin{normalsize}
     \begin{flushright}
     \end{flushright}
    \end{normalsize}
    \begin{LARGE}
       \vspace{1cm}
       \begin{center}
        { A Two Parameter Family\\
               of \\
         the Calabi-Yau d-Fold} \\
       \end{center}
    \end{LARGE}

   \vspace{5mm}

\begin{center}
           Katsuyuki S{\sc ugiyama}\footnote{E-mail address:
             sugiyama@danjuro.phys.s.u-tokyo.ac.jp} \\
       \vspace{4mm}
                  {\it Department of Physics, University of Tokyo} \\
                  {\it Bunkyo-ku, Tokyo 113, Japan} \\
       \vspace{1cm}

     \begin{large} ABSTRACT \end{large}
\par
  \end{center}
\begin{quote}
 \begin{normalsize}

We study a two parameter family of Calabi-Yau d-fold by means of
mirror symmetry. We construct mirror maps and calculate correlation functions
associated with {\kae} moduli in the original manifold. We find there
are more complicated instanton corrections of these couplings than threefolds,
which is expected to reflect families of instantons with continous parameters.

 \end{normalsize}
\end{quote}

\end{titlepage}
\vfill

\section{Introduction}

\pr
It is a long time since the string theory attracted
the attention as a candidate of the unified theory
of elementary particles and their interactions.
A lot of work has been devoted to the study of these
theories, but it seems to be out of reach to gain fundamental
understanding of them.
One of the most important things is the investigation of
the properties of the manifolds on which the string should
be compactified.
Particularlly the compactifications on Calabi-Yau manifolds have received
much attention {\cite{CHSW,SW,CO}}.
{}From the point of view of the particle physics, cohomology classes
of these Calabi-Yau manifolds correspond to zero-mass fields in the low energy
effective theory and these manifolds play crucial roles in deciding
the features of the string theories.

Originally Calabi-Yau threefolds were introduced to provide six
dimensional inner space to yield a consistent string background.
However it seemed hard to investigate their properties because of
the quantum corrections {\cite{DSWW,DG}}.

Recently a great progress has been achieved in
understanding the properties of
the moduli spaces of Calabi-Yau manifolds by the
discovery of the mirror symmetry {\cite{D,LVW,GP,COGP,EM}}.
Now it has become possible to study the structure of the {\kae}
moduli space of the Calabi-Yau manifold to connect with the complex
moduli space of its mirror manifold.

When one discusses the mirror symmetries, the (complex)
dimension of the Calabi-Yau {\mfds} is
restricted to be three because of the string theoretic
applications {\cite{COGP,M,KT,F,LT,KT2,COFKM,CFKM,BCOFHJQ,LSW,HKTY,HKT}}.
Nevertheless as the mathematical physics applications,
generalized mirror manifold with other dimensions are of
interest {\cite{BV,NS,GMR}}.
Lately one parameter family of the Calabi-Yau $d$-fold
realized as a Fermat hypersurface embeded in a projective
space of dimension $(d+1)$ is studied {\cite{GMR}}.
The $d$-point correlation functions studied there have quantum corrections
and these corrections are expected to correspond to
Chern classes of various parameter spaces for rational curves on Calabi-Yau
manifolds. This suggests the much richer structures in higher dimensional
cases.

The aim of this paper is to construct a two parameter family
of the mirror manifold paired with a Calabi-Yau d-fold and
to investigate their properties in order to throw some new lights
upon structures of the moduli spaces.

Firstly we construct the mirror manifold by the method of the
toric varieties {\cite{O,FU}}.
Namely one constructs a mirror manifold as a zero locus of the Laurent
polynomial in the ambient space using the information on the dual
polyhedron $\dpoly$ {\cite{BV,B1,B2,B3,BC}}.
The deformations of the complex structures lead to the deformations of the
Hodge dtructures. The information of the Hodge structures is encoded
in periods of the manifold.
One can correlate the {\kae} moduli space of the original manifold
with complex moduli space of its mirror manifold.
Using the mirror maps, we identify the d point coupling of the
complex moduli on the mirror manifold with that of the {\kae}
moduli on the original manifold. We can calculate quantum corrections of the
latter with this correspondence.

The paper is organized as follows.
In section $2$, we construct a mirror maifold paired with a Calabi-Yau
d-fold $\amb$ in the toric
language {\cite{BV,B1,B2,B3,BC}}. In section $3$, we introduce Gel'fand,
Kapranov and Zelevinsky equation system {\cite{GKZ}} and
identify the Picard-Fuchs equations for the periods of the algebraic
varieties. We construct mirror maps using solutions of this equation
system. In section $4$, the set of correlation functions associated with
the complex moduli and the {\kae} moduli are calculated. The former
correlation functions are meromorphic functions with respect to the
parameters of the manifolds and their singularities are characterized
of the discriminant loci.
By contrast, the latter correlation functions have quantum corrections.
Also we study the {\kae} cone of the original maifold.
In section $5$, monodromy matrices associated with the singular loci
are obtained. Section $6$ is devoted to conclusions and comments.
\\\\
\section{Construction of the mirror \mfd}

\pr
In this section, we consider a Calabi-Yau d-fold $\mm$
defined by a polynomial in the weighted projective space $\amb$
and construct its mirror {\mfd} $\mw$ in the recipe of
the toric variety {\cite{O,FU,BV,B1,B2,B3,BC}}.

\subsection[Construction of the mirror \mfd ]{Polyhedron}

\pr
We choose a lattice polyhedron $\poly$ to construct an ambient space.
The Calabi-Yau manifold is realized by a hypersurface in
this ambient space {\cite{O,FU,B2}}.
The lattice polyhedron $\poly$ is a convex hull $d+1$ dimensinal
rational polyhedron with vertices ${\nu}_i$ \hspace{0.7mm} $(i=1,2,\cdots
,d+2)$
in a lattice space $M$,
\beqs
&&{\nu}_1 :=(d,-1,-1,-1,\cdots ,-1,-1)\,\,\,,\\
&&{\nu}_2 :=(-1,d,-1,-1,\cdots ,-1,-1)\,\,\,,\\
&&{\nu}_3 :=(-1,-1,d,-1,\cdots ,-1,-1)\,\,\,,\\
&&{\nu}_4 :=(-1,-1,-1,d,\cdots ,-1,-1)\,\,\,,\\
&&\cdots \\
&&{\nu}_d :=(-1,-1,-1,-1,\cdots ,d,-1)\,\,\,,\\
&&{\nu}_{d+1} :=(-1,-1,-1,-1,\cdots-1,2d+1)\,\,\,,\\
&&{\nu}_{d+2} :=(-1,-1,-1,-1,\cdots ,-1,-1)\,\,\,.
\eeqs
For each $l$-dimensional face ${\Xi}_l \subset \poly$,
a $d+1$ dimensional cone
$\sigma ({{\Xi}_l})$ is defined as,
\[
\sigma ({{\Xi}_l}):=\{{\lambda \cdot (p-p')
\,\,\,; \,\,\, \lambda \in {\bf R}_{+}\,\,,
p \in \poly \,\,,p' \in {\Xi}_l }\}\,\,\,.
\]
We associate to $\poly$ a complete rational fan $\Sigma ({\poly})$, which
is defined as the collection of $(d+1-l)$ dimensional dual cones
${\sigma}^{\ast} ({{\Xi}_l})$ \hspace{0.5mm} $(l=0,1, \cdots ,d+1)$.
We use a toric variety ${\bf P}_{\poly}$ associated to this fan
$\Sigma ({\poly})$ as the ambient space in order to construct a
Calabi-Yau {\mfd}.
The Calabi-Yau d-fold ${\mm} \equiv {\cal F}({\poly})$ is
realized as a zero locus ${Z_f}$ of the Laurent polynomial $f_{\poly}$
in the algebraic torus $({\bf C}^{\ast})^{d+1} \subset {\bf P}_{\poly}$,
\beq
f_{\poly} (a,X) &:=& \sum^{s}_{i=0} {a_i}{X^{{\nu}_i}} \nonumber \\
&=& {a_0} \nonumber \\
&&+{a_1} {X^d_1}{X^{-1}_2}{X^{-1}_3}\cdots {X^{-1}_d}{X^{-1}_{d+1}}
+{a_2} {X^{-1}_1}{X^{d}_2}{X^{-1}_3}\cdots {X^{-1}_d}{X^{-1}_{d+1}}\nonumber \\
&&+ {a_3} {X^{-1}_1}{X^{-1}_2}{X^{d}_3}\cdots {X^{-1}_d}{X^{-1}_{d+1}}+\cdots
\nonumber \\
&&+{a_{d-1}} {X^{-1}_1}{X^{-1}_2}\cdots {X^{d}_{d-1}}
{X^{-1}_d}{X^{-1}_{d+1}}\nonumber \\
&&+ {a_d} {X^{-1}_1}{X^{-1}_2}\cdots {X^{-1}_{d-1}}{X^{d}_d}{X^{-1}_{d+1}}
+{a_{d+1}} {X^{-1}_1}{X^{-1}_2}{X^{-1}_3}\cdots {X^{-1}_d}
{X^{2d+1}_{d+1}}\nonumber \\
&&+ {a_{d+2}} {X^{-1}_1}{X^{-1}_2}{X^{-1}_3}\cdots {X^{-1}_d}{X^{-1}_{d+1}}
+ \cdots \,\,\,,
\eeq
where the symbol ``$\cdots$'' means
that there are more monomial terms associated
with the deformation of the complex structure.
The toric variety ${\bf P}_{\poly}$ can be described,
\beqs
{\bf P}_{\poly} &:=& \{ \hspace{1mm}
{[{U_0},{U_1},{U_2}, \cdots {U_{d+1}},{U_{d+2}}] \in {\bf P}^{d+3} \,\,;\,\,
{U^2_1}{U^2_2}{U^2_3}\cdots {U^2_{d-1}}{U^2_d}\
{U_{d+1}}{U_{d+2}}={U^{2(d+1)}_0} }\}\,\,\,,\\
{U_1}/{U_0} &=& {X^{d}_1}{X^{-1}_2}{X^{-1}_3}\cdots {X^{-1}_d}{X^{-1}_{d+1}}
\,\,\,,\\
{U_2}/{U_0} &=& {X^{-1}_1}{X^{d}_2}{X^{-1}_3}\cdots {X^{-1}_d}{X^{-1}_{d+1}}
\,\,\,,\\
{U_3}/{U_0} &=& {X^{-1}_1}{X^{-1}_2}{X^{d}_3}\cdots {X^{-1}_d}{X^{-1}_{d+1}}
\,\,\,,\\
\cdots \\
{U_{d-1}}/{U_0} &=& {X^{-1}_1}{X^{-1}_2}{X^{-1}_3}\cdots
{X^{d}_{d-1}}{X^{-1}_d}{X^{-1}_{d+1}} \,\,\,,\\
{U_d}/{U_0} &=& {X^{-1}_1}{X^{-1}_2}{X^{-1}_3}\cdots
{X^{-1}_{d-1}}{X^{d}_d}{X^{-1}_{d+1}} \,\,\,,\\
{U_{d+1}}/{U_0} &=& {X^{-1}_1}{X^{-1}_2}{X^{-1}_3}\cdots
{X^{-1}_{d-1}}{X^{-1}_d}{X^{2d+1}_{d+1}} \,\,\,,\\
{U_{d+2}}/{U_0} &=& {X^{-1}_1}{X^{-1}_2}{X^{-1}_3}\cdots
{X^{-1}_{d-1}}{X^{-1}_d}{X^{-1}_{d+1}} \,\,\,.
\eeqs
Considering the mapping,
\beqs
{X_1}= \frac{z_1}{z_{d+2}^2}\,\,,\,\,{X_2}= \frac{z_2}{z_{d+2}^2}\,\,,
{X_3}= \frac{z_3}{z_{d+2}^2}\,\,,\,\,\cdots \,\,,\,\,
{X_d}= \frac{z_d}{z_{d+2}^2}\,\,,\,\,{X_{d+1}}=
\frac{z_{d+1}}{z_{d+2}}\,\,,\,\,
\eeqs
we can rewrite the zero locus $Z_{f_{\poly}}$
\beqs
&&{a_1}{z^{d+1}_1}+{a_2}{z^{d+1}_2}+\cdots
+{a_{d-1}}{z^{d+1}_{d-1}}+{a_d}{z^{d+1}_d}\\
&&+{a_{d+1}}{z^{2(d+1)}_{d+1}}
+{a_{d+2}}{z^{2(d+1)}_{d+2}}+{a_0}{z_1}{z_2}\cdots {z_{d-1}}
{z_d}{z_{d+1}}{z_{d+2}}+ \cdots =0\,\,\,.
\eeqs
This polynomial in the weighted projective space $\amb$ determines
a Calabi-Yau manifold ${\cal F}({\poly})$ {\cite{B2}}.

\subsection[Construction of the mirror \mfd ]{Mirror manifold}

\pr
A mirror variety ${\mw \equiv }$ ${\cal F}({\dpoly})$ of the manifold
$\mm \equiv$ ${\cal F}({\poly})$ can be constructed by using a dual
polyhedron $\dpoly$ of $\poly$ {\cite{B2}}.
The dual polyhedron $\dpoly$ defined by
\beqs
\dpoly :=\{{({x_1},{x_2},\cdots ,{x_d},{x_{d+1}}) \,\,;\,\,
\sum^{d+1}_{i=1}{x_i}{y_i} \geq -1 \mbox{ for all }
({y_1},{y_2},\cdots ,{y_d},{y_{d+1}}) \in \poly }\} \,\,,
\eeqs
is again a rational polyhedron with vertices ${\nu}^{\ast}_i$
\hspace{0.7mm} $(i=1,2,\cdots ,{d+2})$
in the lattice space $N$,
\beqs
&&\dnu{1} :=(1,0,0,\cdots ,0,0,0) \,\,\,,\\
&&\dnu{2} :=(0,1,0,\cdots ,0,0,0) \,\,\,,\\
&&\dnu{3} :=(0,0,1,\cdots ,0,0,0) \,\,\,,\\
&&\dnu{4} :=(0,0,0,\cdots ,1,0,0) \,\,\,,\\
&&\cdots \\
&&\dnu{d} :=(0,0,0,\cdots ,0,1,0) \,\,\,,\\
&&\dnu{d+1} :=(0,0,0,\cdots ,0,0,1) \,\,\,,\\
&&\dnu{d+2} :=(-2,-2,-2,\cdots ,-2,-2,-1) \,\,\,.
\eeqs
Both the polyhedron $\poly$ and the dual polyhedron $\dpoly$
are rational polyhedrons.
Inside the $\dpoly$, there are two integral points which describe the
exceptional divisors {\cite{OP,BFS,AGM1,AGM2}},
\beqs
&&\dnu{d+3} :=(-1,-1,-1,\cdots ,-1,-1,0) \,\,\,,\\
&&\dnu{0} :=(0,0,0,\cdots ,0,0,0) \,\,\,.
\eeqs
In the same way as in the previous subsection, we construct a Calabi-Yau
variety $\mw \equiv$ ${\cal F}({\dpoly})$ associated to the polyhedron
$\dpoly$. The zero locus of the Laurent polynomial ${f_{\dpoly}}(b,Y)$
in the algebraic torus $({\cal C}^{\ast})^{d+1} \subset {\bf P}_{\dpoly}$
is defined,
\beq
f_{\dpoly} (b,Y) &:=& \sum^{d+3}_{i=0} {b_i}{Y^{\dnu{i}}} \nonumber \\
&=& {b_0}+{b_1}{Y_1}+{b_2}{Y_2}+ \cdots +{b_{d+2}}{Y_{d+2}}+{b_{d+3}}{Y_{d+3}}
\,\,\,.
\eeq
The toric variety ${\bf P}_{\dpoly}$ defined by,
\beqs
{\bf P}_{\dpoly} &:=& \{ \hspace{1mm}
{[{V_0},{V_1},{V_2}, \cdots {V_{d+1}},{V_{d+2}}]
\in {\bf P}^{d+3} \,\,;\,\,{V^2_1}{V^2_2}{V^2_3}\cdots {V^2_d}
{V_{d+1}}{V_{d+2}}={V^{2(d+1)}_0} }\} \,\,\,,\\
{V_1}/{V_0}&=&{Y_1} \,\,,\,\,{V_2}/{V_0} = {Y_2} \,\,,\,\, {V_3}/{V_0} = {Y_3}
\,\,,\cdots \,\,,
{V_{d-1}}/{V_0} = {Y_{d-1}} \,\,,\,\,{V_d}/{V_0} = {Y_d} \,\,,\,\,\\
{V_{d+1}}/{V_0}&=&{Y_{d+1}} \,\,,\,\,
{V_{d+2}}/{V_0} = \frac{1}{{Y^2_1}{Y^2_2}\cdots {Y^2_{d-1}}{Y^2_d}{Y_{d+1}}}
\,\,,\,\,
{V_{d+3}}/{V_0} = \frac{1}{{Y_1}{Y_2}\cdots {Y_{d-1}}{Y_d}} \,\,\,.
\eeqs
is used for the ambient space. Let us introduce an etale map
$\varphi \,\,:\,\, {\bf P}^{d+1} \rightarrow {\bf P}_{\poly}$ defined by,
\beqs
\varphi \,\,&;& [{z_1},{z_2},{z_3},\cdots ,{z_d},{z_{d+1}},{z_{d+2}}] \\
&& \rightarrow [{z_1}{z_2}{z_3}\cdots {z_d}{z_{d+1}}{z_{d+2}},{z^{d+1}_1},
{z^{d+1}_2},{z^{d+1}_3},\cdots ,{z^{d+1}_{d-1}},
{z^{d+1}_d},{z^{2(d+1)}_{d+1}},{z^{2(d+1)}_{d+2}}]\,\,\,.
\eeqs
Then the zero locus $Z_{f_{\dpoly}}$ can be rewritten,
\beq
&&{b_1}{z^{d+1}_1}+{b_2}{z^{d+1}_2}+\cdots +{b_{d-1}}{z^{d+1}_{d-1}}
+{b_d}{z^{d+1}_d}
+{b_{d+1}}{z^{2(d+1)}_{d+1}}+{b_{d+2}}{z^{2(d+1)}_{d+2}} \nonumber\\
&& +{b_{d+3}}{z^{d+1}_{d+1}}{z^{d+1}_{d+2}}+{b_0}{z_1}{z_2}{z_3}\cdots
{z_d}{z_{d+1}}{z_{d+1}} =0
\label{eqn:ZERO1}\,\,\,.
\eeq
The set $({\poly},M)$ is a reflexive pair and the set $({\dpoly},N)$ is a
dual reflexive pair. The lattice polyhedron $\poly$ is a Fano polyhedron
and determines the Gorenstein toric Fano variety
${\bf P}_{\poly}$ {\cite{O,FU,B2}}.
Similarly the dual polyhedron $\dpoly$ determines the dual
Gorenstein toric Fano variety ${\bp}_{\dpoly}$. A zero locus $Z_{f_{\poly}}$
defined by a Laurent polyhedron $f_{\poly}$ with a fixed Newton polyhedron
$\poly$ gives a Calabi-Yau variety ${\cal F}({\poly})$.
A Laurent polynomial $f_{\dpoly}$ with a fixed Newton polyhedron $\dpoly$
leads to another Calabi-Yau variety ${\cal F}({\dpoly})$.
Thus the mirror manifold ${\cal F}({\dpoly})$ of the ${\cal F}({\poly})$
is realized as the zero locus in the
weighted projective space $\amb$ {\cite{B2}}.
In the following sections, we investigate various properties
of this mirror manifold.

\section{Picard-Fuchs equation and the mirror map}

\pr
In this section, we introduce periods of the mirror manifold and write down
a differential equation satisfied by the periods. We solve solutions of
this equation and define the mirror maps of the Calabi-Yau d-fold as the
ratios of the periods. These maps connect the complex structure of
${\cal F}({\dpoly})$ with the {\kae} structure of ${\cal F}({\poly})$.

\subsection[Picard-Fuchs equation and the mirror map]{Periods}

\pr
The structure of the moduli space of a Calabi-Yau manifold is described
by giving the periods of the manifold. The period integrals of the
${\cal F}({\dpoly})$ are defined as {\cite{BCOFHJQ,B1,HKTY,HKT}},
\beq
{{\Pi}_{i}}(b):= \int_{\Gamma_i} \frac{1}{{f_{\dpoly}}(b,X)}
\prod^{d+1}_{j=1} \frac{d {X}_j}{X_j} \label{eqn:PER}\,\,\,,
\eeq
with the homology cycles $\Gamma_i
\in H_{d+1} (({\bc}^{\ast})^{d+1} \verb+\+ Z_{f_{\dpoly}},{\bz})$.
If we choose the cycle $\Gamma$,
\[
\Gamma :=\{\hspace{1mm}{{X_1},{X_2},\cdots ,{X_d},{X_{d+1}}
\in {\bc}^{d+1} \,\,
;\,\, |{X_i}|=1 \,\,(i=1,2,\cdots ,{d+1})}\}\,\,,
\]
the integral is calculated explicitly and the fundamental period
${{\Pi}_0}(b)=\frac{1}{b_0} \vvp{0} (x,y)$ is obtained,
\beq
&&\vvp{0} (x,y)=\sum^{\infty}_{n=0} \sum^{\infty}_{m \geq 2n}
\frac{((d+1) m)!}{(m-2n)!{(m!)^{d}}{(n!)^2}} {x^m}{y^n} \label{eqn:FUN}\,\,\,\\
&& x:=\frac{{b_1}{b_2}{b_3}\cdots {b_{d-1}}{b_d}{b_{d+3}}}
{{b_0}^{d+1}}\,\,,\,\,
y:=\frac{{b_{d+1}}{b_{d+2}}}{{b_{d+3}}^2}\,\,\,.
\eeq
In the next subsection, we derive a differential equation satisfied by
the periods. We solve this equation to determine
the periods instead of carrying out the integral
on the homology cycles explicitly.

\subsection[Picard-Fuchs equation and the mirror map]
{Generalized hypergeometric equation and the Picard-Fuchs equation}

\pr
We will give the Picard-Fuchs equation satisfied by the period integral
of the mirror manifold $\mw \equiv$ ${\cf}_{\dpoly}$
in an efficient way {\cite{B1}}.
Let us introduce the generalized hypergeometric system of Gel'fand, Kapranov
and Zelevinsky {\cite{GKZ}} (GKZ equation) defined by a set of points
$\{{\dnu{0},\dnu{1},\cdots ,\dnu{d+2},\dnu{d+3}}\}$ in the polyhedron $\dpoly$.
Points $\dnu{1},\dnu{2},\cdots ,\dnu{d+1},\dnu{d+2}$ lie at the vertices of
$\dpoly$. On the other hand, points $\dnu{d+3},\dnu{0}$ lie in the
interior of codimension $d$, zero faces of $\dpoly$ respectively.
We embed these integral points in ${\bf R}^{d+2}$ by
\[
\bdnu{i} :=(1,\dnu{i}) \,\,\,(i=0,1,2,\cdots , {d+2},{d+3})\,\,\,.
\]
Then these $(d+4)$ points are linear dependent in ${\bf R}^{d+2}$ and satisfy
two linear relations,
\beqs
&&-(d+1) {\bdnu{0}}+{\bdnu{1}}+{\bdnu{2}}+\cdots +{\bdnu{d-1}}
+{\bdnu{d}}+{\bdnu{d+3}}=0\,\,\,,\\
&&{\bdnu{d+1}}+{\bdnu{d+2}}-2 {\bdnu{d+3}}=0\,\,\,.
\eeqs
We associate two lattice vectors $l^{(1)},l^{(2)}$ in ${\bz}^{d+4}$
to these linear relations,
\beq
l^{(1)} :=(-(d+1),1,1,1,\cdots ,1,1,0,0,1) \label{eqn:L1}\,\,\,,\\
l^{(2)} :=(0,0,0,0,\cdots ,0,0,1,1,-2) \label{eqn:L2}\,\,\,.
\eeq
Using the above things, we define differential operators,
\beq
&&{\BX{i}}:=\prod_{l^{(i)}_j >0}
{\left({\frac{\del}{\del {b_j}}}\right)}^{l^{(i)}_j}
-\prod_{l^{(i)}_j <0}
{\left({\frac{\del}{\del {b_j}}}\right)}^{- l^{(i)}_j}\,\,\,,\\
&&{\cal P}:= \sum^{d+3}_{i=0} \bdnu{i} {b_i} \frac{\del}{\del {b_i}}
-\beta \,\,\,,\\
&&\beta :=(-1,0,0,0,\cdots ,0,0,0,0,0)\,\,\,.\nonumber
\eeq
Then the GKZ differential equation system is defined {\cite{GKZ,HKTY,HKT}},
\beq
&&{\BX{1}} \Phi =0 \label{eqn:GKZ1}\,\,\,\\
&&{\BX{2}} \Phi =0 \label{eqn:GKZ2}\,\,\,\\
&&{\cal P} \Phi =0 \label{eqn:GKZ3}\,\,\,.
\eeq
Explicitly these can be re-expressed as ,
\beq
&&\left\{{{\frac{\del}{\del {b_1}}}{\frac{\del}{\del {b_2}}}\cdots
{\frac{\del}{\del {b_{d-1}}}}
{\frac{\del}{\del {b_d}}}{\frac{\del}{\del {b_{d+3}}}}-
{\left({\frac{\del}{\del {b_1}}}\right)^{d+1}}}\right\} \Phi =0
\label{eqn:HYP1}\,\,\,,\\
&&\left\{{{\frac{\del}{\del {b_{d+1}}}}{\frac{\del}{\del {b_{d+2}}}}-
{\left({\frac{\del}{\del {b_{d+3}}}}\right)^2}}\right\} \Phi =0
\label{eqn:HYP2}\,\,\,,
\eeq
and are also called $\dpoly$-hypergeometric system together {\cite{GKZ,B2}}
with the equation (\ref{eqn:GKZ3}). In order to solve this system, we introduce
the variables,
\beqs
&& x:=\frac{{b_1}{b_2}\cdots {b_{d-1}}{b_d}{b_{d+3}}}{{b_0}^{d+1}}\,\,,\,\,
y:=\frac{{b_{d+1}}{b_{d+2}}}{{b_{d+3}}^2}\,\,\,,\\
&& \vpi (x,y):={b_0} \cdot \Phi (x,y)\,\,\,.
\eeqs
Then the equation (\ref{eqn:GKZ3}) is satisfied automatically,
and the equations (\ref{eqn:HYP1},\ref{eqn:HYP2}) are
reduced to the following equation system,
\beq
&&\th{x}{\cal D}_{l^{(1)}} \vpi (x,y)=0 \nonumber\,\,\,,\\
&&{\cal D}_{l^{(2)}} \vpi (x,y)=0 \nonumber\,\,\,,\\
&&{\cal D}_{l^{(1)}} :={\th{x}}^{N-2} ({\th{x} -2 \th{y}})\\
&&\hspace{2cm} -Nx(N\th{x} +N-1)(N\th{x} +N-2)\cdots (N\th{x} +2)(N\th{x} +1)
\label{eqn:DO1}\,\,\,,\\
&&{\cal D}_{l^{(2)}} :={\th{y}}^2 -y({\th{x} -2\th{y}})
({\th{x} -2\th{y} -1}) \label{eqn:DO2}\,\,\,,
\eeq
where $\th{x} :=x \frac{d}{dx}$ ,$\th{y} :=y \frac{d}{dy}$ and $N:=d+1$.
In order to associate the solutions in the above system to the periods
(\ref{eqn:PER})
defined in the section $3.1$, we take the fundamental period
${\Pi}_0 =\frac{1}{b_0} \vvp{0}$. As is seen easily,
$\vvp{0}$ (\ref{eqn:FUN}) satisfies
an equation system,
\beq
&&{\cal D}_{l^{(1)}} \vvp{0} (x,y)=0\,\,\,,\\
&&{\cal D}_{l^{(2)}} \vvp{0} (x,y)=0\,\,\,.
\eeq
Furthermore the periods ${\Pi}_0$ satisfys the equation,
\beq
{\cal P}{\Pi}_0 =0\,\,\,.
\eeq
Taking into account of this fact, we make the Ansatz that
the period integrals ${\Pi}(b)$ satisfy the above
$\dpoly$-hypergeometric system and are identified with the solutions
$\Phi (b)$ except that the ${b_0} \Pi (b)$ is included in the
Ker${{\cal D}_{l^{(1)}}}$ rather than Ker$\{{\th{x} {\cal D}_{l^{(1)}}}\}$.
Under this Ansatz, we identify ${b_0}\Pi (b)$
with the solution $\vpi (x,y)$ of the simultaneous differential equations,
\beq
&&{\cal D}_{l^{(1)}} \vpi (x,y)=0 \label{eqn:SDE1}\,\,\,,\\
&&{\cal D}_{l^{(2)}} \vpi (x,y)=0 \label{eqn:SDE2}\,\,\,.
\eeq
and identify the Picard-Fuchs equations for the periods with
the GKZ equation system.

\subsection[Picard-Fuchs equation and the mirror map]{The mirror map}

\pr
In this section, we construct mirror maps $t_1 ,t_2 $ which connect
the {\kae} structure of the original {\mfd} ${\cf}({\poly})$ with the complex
structure of the mirror {\mfd} ${\cf}({\dpoly})$ by using solutions of
the $\dpoly$-hypergeometric system.

The GKZ differential equation system has maximally unipotent
monodromy {\cite{M}}
at a point $(x,y)=(0,0)$. This point $(x,y)=(0,0)$ corresponds to the
large complex structure of the mirror manifold $\mw \equiv$ ${\cf}({\dpoly})$.
We impose the condition that the large complex structure limit of
the mirror manifold ${\cf}({\poly})$ matches the large radius limit of
the original manifold ${\cf}({\poly})$. In order to satisfy this claim,
we takes as a boundary condition of the mirror maps $t_1 ,t_2$
the following conditions,
\beqs
&&{t_1}({e^{2 \pi i}x,y})={t_1}(x,y)+1\,\,\,,\\
&&{t_2}({x,{e^{2 \pi i}}y})={t_2}(x,y)+1\,\,\,.
\eeqs
These transformation are translations and correspond to the
modular transformation at the infinity.
These maps $t_1 ,t_2$ can be expressed by the periods
${\vpi}^{(0)},{\vpi}^{(1)},{\vpi}^{(2)}$, which satisfy the GKZ equation
system. Considering the boundary conditions, we can define the mirror maps
$t_1 , t_2$,
\beq
&&{t_1}(x,y):=\frac{1}{2 \pi i} \frac{{\vpi}^{(1)}}{{\vpi}^{(0)}}
\label{eqn:MIR1}\,\,\,,\\
&&{t_2}(x,y):=\frac{1}{2 \pi i} \frac{{\vpi}^{(2)}}{{\vpi}^{(0)}}
\label{eqn:MIR2}\,\,\,.
\eeq
In this formula, ${\vpi}^{(0)},{\vpi}^{(1)},{\vpi}^{(2)}$ are solutions of
the equations (\ref{eqn:SDE1},\ref{eqn:SDE2}) and are given as,
\beq
{{\vpi}^{(0)}}(x,y)&:=&\sum_{m,n \geq 0}
\frac{\gam{(d+1)m+1}}{{\left[{\gam{m+1}}\right]}^d}
\frac{{x^m}{y^n}}{\gam{m-2n+1} {[{\gam{n+1}}]}^2}\label{eqn:P1}\,\,\,,\\
{{\vpi}^{(1)}}(x,y)&:=&{{\vpi}^{(0)}}(x,y)\cdot \log x \nonumber \\
&+&\sum_{m,n \geq 0}
\frac{\gam{(d+1)m+1}}{{\left[{\gam{m+1}}\right]}^d}
\frac{{x^m}{y^n}}{\gam{m-2n+1} {[{\gam{n+1}}]}^2} \nonumber \\
&\times &\left[{(d+1)\Psi ((d+1)m+1)-d \Psi (m+1)-\Psi(m-2n+1)}\right]
\label{eqn:P2}\,\,\,,\\
{{\vpi}^{(2)}}(x,y)&:=&{{\vpi}^{(0)}}(x,y)\cdot \log y \nonumber \\
&+&\sum_{m,n \geq 0}
\frac{\gam{(d+1)m+1}}{{\left[{\gam{m+1}}\right]}^d}
\frac{{x^m}{y^n}}{\gam{m-2n+1} {[{\gam{n+1}}]}^2} \nonumber \\
&\times &2 \left[{\Psi(m-2n+1)-\Psi (n+1)}\right]\label{eqn:P3}\,\,\,.
\eeq
In the later convenience, we rewrite the maps $t_1 , t_2 $
in the integral representation. Firstly we introduce the parameters
$\psi ,\phi$,
\[
\psi := -\frac{1}{2N} {(x^2 y)}^{-1/(2N)}\,\,,
\,\,\phi :=-\frac{1}{2} y^{-1/2}\,\,\,.
\]
Then the zero locus (\ref{eqn:ZERO1}) becomes the following formula with
a bit of rescaling of the set of variables,
\beq
{\it p}&=&{z^{d+1}_1}+{z^{d+1}_2}+\cdots +{z^{d+1}_{d-1}}+{z^{d+1}_d}
+{z^{2(d+1)}_{d+1}}+{z^{2(d+1)}_{d+2}}\nonumber \\
&&-2(d+1)\psi {z_1}{z_2}\cdots {z_{d-1}}{z_d}{z_{d+1}}{z_{d+2}}-
2\phi {z_{d+1}}^{d+1} {z_{d+2}}^{d+1} =0 \label{eqn:ZERO2}\,\,\,.
\eeq
With this new variables, the series (\ref{eqn:P1}) can be rewritten as
\beq
{{\vpi}^{(0)}}&:=&{{\vpi}^{(0)}}({\psi ,\phi}) \nonumber \\
&=&\sum^{\infty}_{n=0}
\frac{\gam{(d+1)n+1}}{{\left[{\gam{n+1}}\right]}^{d+1}}
\frac{{(-1)}^n}{{(2(d+1)\psi)}^{(d+1)n}}
u_n ({\phi})\,\,\,,\\
{u_{\nu}}(\phi)&:=&{{(2\phi)}^{\nu}}{{}_2}F_1 \left({-\frac{\nu}{2},
-\frac{\nu}{2}+\frac{1}{2},1;\frac{1}{{\phi}^2}}\right) \label{eqn:UN}\,\,\,.\\
&&\mbox{for}\,\,\left|{\frac{2^d {\psi}^{d+1}}{\phi\pm 1}}\right| >1 \,\,
\mbox{and}\,\, -\pi <arg \left({\frac{2^d {\psi}^{d+1}}
{\phi\pm 1}}\right) < \pi \nonumber\,\,\,.
\eeq
By the analytic continuation in the region
\[
\left|{\frac{2^d {\psi}^{d+1}}{\phi\pm 1}}\right| <1 \,\,\,,
\]
${\vvp{0}}(\psi ,\phi)$ can be expressed,
\beq
{\vvp{0}}({\psi ,\phi})=-\frac{1}{(d+1) {\pi}^d} \sum^{\infty}_{m=1}
\frac{{\left[{\gamm{m}{d+1}}\right]}^{d+1}}{\gam{m}} \times
{\left({\sin \frac{\pi m}{d+1}}\right)}^d {{(-2(d+1)\psi)}^m}
{u_{-\frac{m}{d+1}}}({\phi})\,\,\,.
\eeq
A complete set of solutions valid in this range is easily constructed.
The set of solutions $\{{\vvp{j}}\}$\,\,$(j=0,1,2,\cdots , {d+3},{d+4})$
defined by,
\beq
{\vvp{j}}({\psi ,\phi})&=&{\vvp{0}}
({{{\alpha}^j}\psi ,{{(-1)}^j}\phi})\nonumber \\
&=&-\frac{1}{(d+1) {\pi}^d} \sum^{\infty}_{m=1}
\frac{{\left[{\gamm{m}{d+1}}\right]}^{d+1}}{\gam{m}} \nonumber \\
&& \times {\left({\sin \frac{\pi m}{d+1}}\right)}^d
{{(-2(d+1)\psi)}^m} {{\alpha}^{jm}}
{u_{-\frac{m}{d+1}}}({{{(-1)}^j}\phi})\,\,\,,\\
{\alpha}&=&\exp{\frac{2 \pi i}{2(d+1)}}\,\,\,,
\eeq
expands the complete set of solutions.
Integral representations of these solution can be represented as,
\beq
&&{\vvp{2j}}({\psi ,\phi})=\frac{-1}{(d+1) {\pi}^d} \sum^d_{r=1} {(-1)}^r
{\left({\sin \frac{\pi r}{d+1}}\right)}^d \cdot {\alpha}^{2jr}
{\xi}_r ({\psi},{\phi})\,\,\,,\\
&&{\vvp{2j+1}}({\psi ,\phi})=\frac{-1}{(d+1) {\pi}^d} \sum^d_{r=1} {(-1)}^r
{\left({\sin \frac{\pi r}{d+1}}\right)}^d \cdot {\alpha}^{(2j+1)r}
{\eta}_r ({\psi},{\phi})\,\,\,,\\
&&{{\xi}_r}({\psi},{\phi})=\int_C
\frac{d\nu}{2i \sin\pi \left({\nu +\frac{r}{d+1}}\right)} \cdot
\frac{{\left[{\gam{-\nu}}\right]}^{d+1}}{\gam{-(d+1)\nu}} \cdot
{(2(d+1)\psi)}^{-(d+1)\nu}
{u_{\nu}}({\phi})\,\,\,,\\
&&{{\eta}_r}({\psi},{\phi})=\int_C
\frac{-d\nu}{2i \sin\pi \left({\nu +\frac{r}{d+1}}\right)} \cdot
\frac{{\left[{\gam{-\nu}}\right]}^{d+1}}{\gam{-(d+1)\nu}} \cdot
{(2(d+1)\psi)}^{-(d+1)\nu}\nonumber \\
&&\;\;\;\;\;\;\,\,\,\times{\displaystyle
\frac{{u_{\nu}}({\phi})\sin\pi \left({\nu +\frac{r}{d+1}}\right) +
{u_{\nu}}({-\phi}) \sin \frac{\pi r}{d+1}}{\sin\pi\nu}}\,\,\,,
\;\;\;(j=0,1,2,\cdots ,{d-1})\,\,,
\eeq
where the contour $C$ is chosen to enclose the poles
${\displaystyle \nu =-n-\frac{r}{d+1}}$\,\,$(n=0,1,2,\cdots)$ counterclockwise.
With this basis, the mirror maps $t_1 , t_2 $  (\ref{eqn:MIR1},\ref{eqn:MIR2})
can be re-expressed,
\beq
2\pi i{t_1}&=&\frac{-\vvp{0}+(d-1)\vvp{2}+(d-2)\vvp{4}+\cdots
+2\vvp{2(d-2)}+\vvp{2(d-1)}}{(d+1)\vvp{0}}\,\,\,,\\
&=&-\frac{1}{2} \\
&& +\frac{1}{4\vvp{0}} \int_C
\frac{d\nu}{{(\sin\pi\nu)}^2} \cdot \frac{\gam{(d+1)\nu +1}}
{{\left[{\gam{\nu +1}}\right]}^{d+1}} \cdot
{(2(d+1)\psi)}^{-(d+1)\nu} {u_{\nu}}({\phi})\cos\pi\nu \label{eqn:IR1}\,\,\,,\\
2\pi i{t_2}&=&-\frac{-\vvp{0}+(d-1)\vvp{2}+(d-2)\vvp{4}+\cdots
+2\vvp{2(d-2)}+\vvp{2(d-1)}}
{(d+1)\vvp{0}} \nonumber \\
&+&\frac{{d}\vvp{1}+(d-1)\vvp{3}+(d-2)\vvp{5}+\cdots
+2\vvp{2d-3}+\vvp{2d-1}}
{(d+1)\vvp{0}}\,\,\,,\\
&=&\frac{1}{2}-\frac{1}{4\vvp{0}} \int_C
\frac{d\nu}{{(\sin\pi\nu)}^2} \cdot \frac{\gam{(d+1)\nu +1}}{{\left[{\gam{\nu
+1}}\right]}^{d+1}} \cdot {(2(d+1)\psi)}^{-(d+1)\nu} \nonumber \\
&&\times \left[{{u_{\nu}}({\phi})\cos\pi\nu -{u_{\nu}}({-\phi})}\right]
\label{eqn:IR2}\,\,\,.
\eeq
When we take the contour $C$ to enclose the poles
${\displaystyle \nu =-n-\frac{r}{d+1}}$\,\,$(n+0,1,2,\cdots)$ counterclockwise,
the above formulae coincide with the representation
(\ref{eqn:MIR1},\ref{eqn:MIR2}).

\section{The correlation functions}

\pr
In this section, we calculate d-point correlation functions associated
with the complex structure of mirror manifold ${\cf}({\dpoly})$. Also
we investigate the {\kae} cone of the original manifold ${\cf}({\poly})$
and d-point correlation functions of cohomology classes
associated with the {\kae} structure.

\subsection[The correlation functions]
{Correlation functions associated with the complex structure}

\pr
It is known that there exists a nowhere-vanishing holomorphic $d$-form
$\Om$ with respect to a canonical basis
$({\alpha}_i ,{\beta}^j)\in $
$H^d ({({{\bc}^{\ast}})}^{d+1} \verb+\+ Z_{f_{\dpoly}},{\bz})$
\[
\Om =\sum_{i}
({z^i {\alpha}_i - g_i {\beta}^i}) \,\,\,
\]
with
\beqs
&&\int_{\mw} {\alpha}_i \wedge {\alpha}_j
=\int_{\mw} {\beta}^i \wedge {\beta}^j =0\,\,\,,\\
&&\int_{\mw} {\alpha}_i \wedge {\beta}^j ={\delta}^j_i \,\,\,,
\eeqs
The set $({z^i},{g_j})$ are realized as linear combinations of
the periods and can be written as linear combination of the solutions
of the GKZ equation (\ref{eqn:SDE1},\ref{eqn:SDE2}).

Now let us consider a change of the complex structure of the
Calabi-Yau manifold. With the help of the Kodaira~Spencer theorem, the
following relations are understood,
\beqs
\frac{\del\Om}{\del {{\zeta}^{i_1}}}&\in&{H^{(d,0)}}
\otimes H^{({d-1},1)}\,\,\,,\\
\frac{{\del}^2 \Om}{\del {{\zeta}^{i_1}}\del {{\zeta}^{i_2}}} &\in&
H^{(d,0)} \otimes H^{({d-1},1)} \otimes H^{({d-2},2)}\,\,\,,\\
\frac{{\del}^3 \Om}{\del {{\zeta}^{i_11}}\del {{\zeta}^{i_2}}
\del {{\zeta}^{i_3}}} &\in&
H^{(d,0)} \otimes H^{({d-1},1)} \otimes H^{({d-2},2)} \otimes
H^{({d-3},3)}\,\,\,,\\
\cdots \\
\frac{{\del}^{d-1} \Om}{\del {{\zeta}^{i_1}}\del {{\zeta}^{i_2}}
\cdots \del {{\zeta}^{i_{d-1}}}} &\in&
H^{(d,0)} \otimes H^{({d-1},1)} \otimes H^{({d-2},2)} \otimes \cdots \otimes
H^{(2,{d-2})}
\otimes H^{(1,{d-1})}\,\,\,,\\
\frac{{\del}^d \Om}{\del {{\zeta}^{i_1}}\del {{\zeta}^{i_2}}\cdots
\del {{\zeta}^{i_{d-1}}}\del {{\zeta}^{i_{d}}}} &\in&
H^{(d,0)} \otimes H^{({d-1},1)} \otimes H^{({d-2},2)} \otimes \cdots \otimes
H^{(2,{d-2})}
\otimes H^{(1,{d-1})} \otimes H^{(0,d)}\,\,\,.
\eeqs
Clearly the following equations are satisfied,
\beq
&&\int_{\mw} \Om\wedge\frac{\del\Om}{\del {{\zeta}^{i_1}}} =0\,\,,\,\,
\int_{\mw} \Om\wedge\frac{{\del}^2 \Om}{\del {{\zeta}^{i_1}}
\del {{\zeta}^{i_2}}} =0\,\,,
\nonumber\\
&&\int_{\mw} \Om\wedge
\frac{{\del}^3 \Om}{\del {{\zeta}^{i_1}}\del {{\zeta}^{i_2}}\del
{{\zeta}^{i_3}}} =0\,\,,\,\,\cdots \,\,
\int_{\mw} \Om\wedge
\frac{{\del}^{d-1} \Om}
{\del {{\zeta}^{i_1}}\del {{\zeta}^{i_2}}\cdots \del {{\zeta}^{i_{d-2}}}
\del {{\zeta}^{i_{d-1}}}} =0
\label{eqn:KS}\,\,.
\eeq
We can define d-point correlation functions
${\Kappa}_{{i_1}{i_2}\cdots {i_d}}$,
\beq
{{\Kappa}_{{i_1}{i_2}\cdots {i_d}}}&:=&- \int_{\mw} \Om\wedge
\frac{{\del}^d \Om}{\del {{\zeta}^{i_1}}\del {{\zeta}^{i_2}}\del
{{\zeta}^{i_3}}\cdots \del {{\zeta}^{i_{d-1}}}\del {{\zeta}^{i_d}}} \,\,\,\\
&=&\sum_n ({{z^n} {\del}_{i_1} {\del}_{i_2} \cdots {\del}_{i_{d-1}}
{\del}_{i_d} {g_n}
-{g_n} {\del}_{i_1} {\del}_{i_2} \cdots {\del}_{i_{d-1}} {\del}_{i_d}
{z^n}})\,\,\,.
\eeq
Because the sets $({z^n},{g_n})$ can be expressed as linear combinations
of the solutions $\vpi (x,y)$ of the GKZ equation, we can obtain
the couplings ${\Kappa}_{{i_1}{i_2}\cdots {i_d}}$ from the information on the
GKZ
equation. For convenience, we define the following variables,
\beq
&&\bx :=N^N x\,\,,\,\,\by :=4y\,\,,\\
&&\w{a,b} :=\sum_{n} ({{z^n}\,{\del}^a_{\bx}\,{\del}^b_{\by}\,{g_n}
-{g_n}\,{\del}^a_{\bx}\,{\del}^b_{\by}\,{z^n}})\,\,\,,
\eeq
Then the differential operators ${\cal D}_{l^{(1)}}$,
${\cal D}_{l^{(2)}}$ (\ref{eqn:DO1},\ref{eqn:DO2}) can be rewritten as,
\beqs
&&{\cal D}_{l^{(1)}} = {\bx}^{N-1} \,(1-\bx)\,{\del}^{N-1}_{\bx}
-2\,{\bx}^{N-2} \,\by\,
{\del}^{N-2}_{\bx}\,{\del}_{\by} + {\bx}^{N-2}
\,\left({\frac{(N-1)(N-2)}{2}-\frac{{(N-1)}^2}{2}\bx}\right) \,
{\del}^{N-2}_{\bx}\\
&&-(N-2)(N-3)\, {\bx}^{N-3} \,\by\, {\del}^{N-3}_{\bx}\,{\del}_{\by} + \cdots
\,\,\,,\\
&&{\cal D}_{l^{(2)}} = {\by}^2 \,(1-\by)\,{\del}^2_{\by} +
\by\,\left({1-\frac{3}{2} \by}\right) \, {\del}_{\by}
+\bx\,{\by}^2 \,{\del}_{\bx}\,{\del}_{\by} -\frac{1}{4}
{\bx}^2 \,\by\,{\del}^2_{\bx}\,\,\,,
\eeqs
and the GKZ equation system is written,
\beq
{\cal D}_{l^{(1)}} \vpi =0\,\,,\,\,{\cal D}_{l^{(2)}} \vpi =0\,\,.
\eeq
The sets $({z^n},{g_n})$ are linear combinations of
the above solutions $\vpi$.
The $\w{a,b}$ satisfy the following equations,
\beqs
&&\bx\,(1-\bx)\,\w{d,0} -2\,\by\,\w{{d-1},1} =0\,\,\,,\\
&&\by\,(1-\by)\,\w{0,d} +\bx\,\by\,\w{1,{d-1}}-\frac{1}{4}\,{\bx}^2
\,\w{2,{d-2}} =0\,\,\,,\\
&&\by\,(1-\by)\,\w{1,{d-1}} +\bx\,\by\,\w{2,{d-2}}-\frac{1}{4}\,{\bx}^2
\,\w{3,{d-3}} =0\,\,\,,\\
&&\by\,(1-\by)\,\w{2,{d-2}} +\bx\,\by\,\w{3,{d-3}}-\frac{1}{4}\,{\bx}^2
\,\w{{d-1},1} =0\,\,\,,\\
\cdots\\
&&\by\,(1-\by)\,\w{{d-2},2} +\bx\,\by\,\w{{d-1},1}-\frac{1}{4}\,{\bx}^2
\,\w{d,0} =0\,\,\,.
\eeqs
Using several identities derived from (\ref{eqn:KS}),
\beqs
&&\w{N,0} =\frac{N}{2}\,{\del}_{\bx} \,\w{5,0}\,\,\,,\\
&&\w{{N-1},1} =\frac{1}{2} ((N-1)\,{\del}_{\bx} \,\w{{N-2},1}+{\del}_{\by}
\,\w{{N-1},0}) \,\,\,,\\
&&\w{{N-2},2} = \frac{1}{2} ((N-2)\,{\del}_{\bx} \,\w{{N-3},2}+2\,{\del}_{\by}
\,\w{{N-2},1})\,\,\,,
\eeqs
we calculate the d point coupling $\w{a,b}\,\,(a+b=d)$ up to
a common overall constant factor by solving the
simultaneous equation,
\beqs
&&{\bx}^2 \,(1-\bx)\,\w{N,0} -2\,\bx\,\by\,\w{{N-1},1}
+\bx\,\left({\frac{N(N-1)}{2}-\frac{N^2 +1}{2}\,\bx}\right)\,\w{{N-1},0}\\
&&\,\,\, -(N-1)(N-2)\,\by\,\w{{N-2},1} =0\,\,\,,\\
&&\by\,(1-\by)\,\w{{N-2},2} +\left({1+\left({N-\frac{7}{2}}\right)\,
\by}\right)\,\w{{N-2},1}
-\frac{N-2}{2}\,\bx\,\w{{N-1},0} \\
&&\,\,\, +\bx\,\by\,\w{{N-1},1}-\frac{1}{4}\,{\bx}^2 \,\w{N,0}=0\,\,\,.
\eeqs
Thus we obtain the solutions,
\beqs
&&\w{d,0} =\frac{1}{{\poly}_1 {\poly}^d_3}\,\,\,,\\
&&\w{{d-1},1} =\frac{1-\bx}{2 {\poly}_1 {\poly}^{d-1}_3 {\poly}_4}\,\,\,,\\
&&\w{{d-2},2} =\frac{-1+2\bx }{4 {\poly}_1 {\poly}_2
{\poly}^{d-2}_3 {\poly}_4}\,\,\,,\\
&&\w{{d-3},3} =\frac{1-\bx +\by -3\bx\by}{8 {\poly}_1 {\poly}^2_2
{\poly}^{d-3}_3 {\poly}^2_4}\,\,\,,\\
&&\w{{d-4},4} =\frac{-3 + 4\bx -\by +4\bx\by}{16 {\poly}_1 {\poly}^3_2
{\poly}^{d-4}_3 {\poly}^2_4}\,\,\,,\\
&&\w{{d-5},5} =\frac{1-\bx +6\by -10\bx\by + {\by}^2 -5\bx {\by}^2}
{32 {\poly}_1 {\poly}^4_2 {\poly}^{d-5}_3 {\poly}^3_4}\,\,\,.
\eeqs
We obtain general formula\,\,,
\beqs
\w{d-l,l}=\frac{f_l (\bx ,\by)}{2^l {\poly}_1 {\poly}^{l-1}_2
{\poly}^{d-l}_3 {\poly}^{[{\frac{l+1}{2}}]}_4}\,\,,\,\,(l=1,2,\cdots ,d)\,\,.
\eeqs
where
\beqs
&&f_{2m+1} = \frac{1}{2}\sum^{m-1}_{l=0} {{}_{2m}}{C_{2l+1}} \cdot {\by}^l
\cdot f_3
-{(1-{\by})}^2
\sum^{m-2}_{l=0} {{}_{2(m-1)}}{C_{2l+1}} \cdot {\by}^l \cdot f_1 \,\,\,,\\
&&f_{2m+2} = \frac{1}{2}\sum^{m-1}_{l=0} {{}_{2m}}{C_{2l+1}} \cdot {\by}^l
\cdot f_4
-{(1-{\by})}^2
\sum^{m-2}_{l=0} {{}_{2(m-1)}}{C_{2l+1}} \cdot {\by}^l \cdot f_2 \,\,\,,\\
&&f_1 :=1-\bx \,\,\,,\,\, f_2 :=-1+2 \bx \,\,\,,\\
&&f_3 :=1-\bx +\by -3 \bx \by \,\,\,,\,\, f_4 :=-3+4 \bx -\by +4\bx \by
\,\,\,,\\
&&\hspace{3cm}(m=2,3,4, \cdots)\,\,\,.
\eeqs

Here the ${\poly}_1 \,,\,{\poly}_2 \,,\,{\poly}_3 \,,\,{\poly}_4 $
are discriminants of the differential equation and are defined as,
\beqs
&&{\poly}_1 = {(1-\bx)}^2 -{\bx}^2 \,\by\,\,\,,\\
&&{\poly}_2 = 1-\by \,\,\,,\\
&&{\poly}_3 = \bx\,\,\,,\\
&&{\poly}_4 = \by\,\,\,.
\eeqs
The above couplings get singular when these discriminant loci vanish.
In short, the singular properties of the correlation function associated
with the complex structure of the ${\cf}({\dpoly})$ are encoded
in the discriminant loci of the $\dpoly$-hypergeometric system.
Using this formula, we can calculate {\kae} couplings $K_{a,b}$.
\beqs
K_{a,b}:=\frac{1}{{\varpi}^2_0}\sum_{l+m=a,m+r=b,l+m=c,n+r=d}
\w{c,d} \cdot {\left({\frac{dx}{dt_1}}\right)}^l \cdot
{\left({\frac{dx}{dt_2}}\right)}^m \cdot
{\left({\frac{dy}{dt_1}}\right)}^n \cdot
{\left({\frac{dy}{dt_2}}\right)}^r {{}_a}C_l  \cdot {{}_b}C_m \,\,\,.
\eeqs
We obtain the following results up to degree one,
\beqs
&&K_{d,0}:=2N+2N \cdot \Big\{{2\cdot N^N -(N+1)\cdot N!-(N-1)\cdot N!
\left({\sum^N_{m=2} \frac{N}{m}}\right)}\Big\} q_1 +\cdots \,\,\,,\\
&&K_{d,0}:=N+N \cdot \Big\{{N^N -2\cdot N!-(N-2)\cdot N!
\left({\sum^N_{m=2} \frac{N}{m}}\right)}\Big\} q_1 +\cdots \,\,\,,\\
&&K_{d,0}:=0+\cdots \,\,\,,\\
&&K_{d,0}:=N q_2 +\cdots \,\,\,,\\
&&K_{{d-n},n}:=0+\cdots \,\,\,,\,\,(n \geq 4)\,\,,\\
&&\hspace{3cm} q_1 :=\exp (2\pi i t_1)\,\,\,\,\,
q_2 :=\exp (2\pi i t_2)\,\,\,\,\,.
\eeqs
\subsection[The correlation functions]{The {\kae} cone}

\pr
In this subsection, we will be concerned with the {\kae} moduli space
of the original {\mfd} ${\cf}({\poly})$. This space has the structure
of a cone, called the {\kae} cone in contrast with the structure of the
complex complex moduli space of the ${\cf}({\dpoly})$.
The {\kae} cone is defined by the requirement of a {\kae} form on the
Calabi-Yau {\mfd} ${\cf}({\poly})$,
\beqs
&&\int_{\mm} K \wedge K \wedge K \wedge\cdots\wedge K \wedge K >0\,\,,\,\,
\int_{S({d-1})} K \wedge K \wedge\cdots\wedge K \wedge K >0\,\,,\\
&&\int_{S({d-2})} K \wedge K \wedge\cdots\wedge K  >0\,\,,\cdots \,\,,
\int_{S(2)} K \wedge K  >0\,\,,\,\,
\int_{S(1)} K >0\,\,,
\eeqs
where $S(l)\,(l=1,2,\cdots ,{d-1})$ means the $l$-dimensional homologically
nontrivial hypersurfaces in ${\cf}({\poly})$.
The above requirement are translated into the toric language
{\cite{OP,HKTY,HKT}}.
Firstly let us consider the new polyhedron
${\bar{\Delta}}^{\ast} :=(1,\dpoly)$ in
${\bf R}^{d+2}$. We decompose the polyhedron ${\bar{\Delta}}^{\ast}$ into
simplices. When we take a $(d+2)$ dimensional simplex $\sigma$,
an arbitrary point $v \in \sigma $ can be written as,
\beqs
&&v= {c_{i_0}}{\bdnu{i_0}}+{c_{i_1}}{\bdnu{i_1}}
+\cdots +{c_{i_N}}{\bdnu{i_N}}\,\,\,,\\
&&\mbox{with}\,\,\,{c_{i_0}}+{c_{i_1}}+\cdots +{c_{i_N}} \leq 1 \,\,,
{c_{i_0}}>0\,,\,{c_{i_1}}>0\,,  \cdots ,\,{c_{i_N}}>0\,,
\eeqs
where ${\bdnu{i_0}},{\bdnu{i_1}},\cdots,{\bdnu{i_N}}$ are
the vertices of the $\sigma$ and the subset of the
$\{{{\bdnu{0}},{\bdnu{1}},\cdots,{\bdnu{N+2}}}\}$.
A piecewise linear function $u$ is determined by giving a real value
$u_i \in {\bf R}$ for each vertex ${\bdnu{i}}\in {\bar{\Delta}}^{\ast} $.
Then the piecewise linear function $u$ takes the value for the point
$v\in \sigma$,
\[
u(v)={c_{i_0}}{u_{i_0}}+{c_{i_1}}{u_{i_1}}+
\cdots +{c_{i_N}}{u_{i_N}}\,\,\,.
\]
Equivalently, one can define this piecewise linear function by associating a
vector $z_{\sigma} \in {\bf R}^{N+1} $ to each $(d+2)$ dimensional simplex
$\sigma$,
\[
u(v):=\ipro{z_{\sigma},v}\,\,\,\mbox{for all}\,\,v\in \sigma\,\,,
\]
where $\ipro{{\ast},{\ast}}$ is the Euclidean inner product.
Out of the piecewise linear functions, strictly
convex piecewise linear functions defined by,
\beqs
&&u(v)=\ipro{z_{\sigma} ,v}\,\,\,\mbox{when}\,\,v\in \sigma \,\,\,,\\
&&u(v)>\ipro{z_{\sigma} ,v}\,\,\,\mbox{when}\,\,v {\not\in }\sigma \,\,\,,
\eeqs
play an important role.
The strictly convex piecewise linear functions constitute a cone in the
following quotient space $V$,
\[
V:= \{{\sum^{d+3}_{i=0} {\bf R} \,\be{i}}\}/\sim \,\,\,,
\]
where the symbol "$\sim$" means the relations between the vectors $\be{i}$,
\beq
\sim \,\,;\,\,\sum^{d+3}_{i=0} \ipro{x,\bdnu{i}}\,\be{i}=0
\,\,\,\mbox{for all}\,\,x\in {\bf R}^{d+1} \label{eqn:REL1}\,\,\,.
\eeq

In the context of the toric variety, this cone can be identified with the
{\kae} cone of ${\bp}_{\poly}$ {\cite{O,B2,B3,HKTY}}.
By means of this method, we can get the
{\kae} cone of ${\cf}({\poly})$.
If we take a $(d+2)$ dimensional simplex $\sigma$,
\[
\sigma = \ipro{0,\bdnu{0},\bdnu{2},\bdnu{3},
\bdnu{4},\cdots ,\bdnu{d+1},\bdnu{d+2}}\,\,\,,
\]
then the vector $z_{\sigma}$ can be written,
\beqs
{z_{\sigma}}&=&({u_0},d{u_0}-{u_2}-\cdots -{u_{d-1}}-{u_d}-\frac{1}{2}
{u_{d+1}}-\frac{1}{2}{u_{d+2}},\\
&&-{u_0}+{u_2},-{u_0}+{u_3},
-{u_0}+{u_4},\cdots ,-{u_0}+{u_{d+1}})\,\,\,.
\eeqs
The conditions of the strict convexity on $u$ reads
\beqs
&&-2(d+1){u_0}+2({u_1}+{u_2}+\cdots
+{u_{d-1}}+{u_d})+{u_{d+1}}+{u_{d+2}}>0\,\,\,,\\
&&2{u_{d+3}}-{u_{d+1}}-{u_{d+2}}>0\,\,\,.
\eeqs
Also the relations (\ref{eqn:REL1}) are written down explicitly,
\beqs
&&\be{0}+\be{1}+\be{2}+\cdots
+\be{d-1}+\be{d}+\be{d+1}+\be{d+2}+\be{d+3}=0\,\,\,,\\
&&\be{1}=\be{2}=\cdots =\be{d-1}=\be{d}=2\be{d+2}+\be{d+3}\,\,\,,\\
&&\be{d+1}=\be{d+2}\,\,\,.
\eeqs
Then the generic element $K_u$ in $V$ can be expressed as,
\beq
{K_u}&=&-\sum^{d+3}_{i=0} u_i \be{i}\nonumber \\
&=&\frac{1}{2(d+1)}[{-2(d+1){u_0}+2({{u_1}+{u_2}+\cdots +{u_{d-1}}+{u_d}})
+{u_{d+1}}+{u_{d+2}}}] \,\be{0}
\nonumber \\
&+&\frac{1}{2}[{{u_{d+1}}+{u_{d+2}}-2{u_{d+3}}}] \,\be{d+3}\nonumber \\
&=&\frac{1}{2(d+1)}\ipro{u,2 l^{(1)} + l^{(2)}} \,\be{0} +
\frac{1}{2}\ipro{u, l^{(2)}} \,\be{d+3}\,\,\,,
\eeq
with the conditions,
\beqs
&&\ipro{u,2 l^{(1)} + l^{(2)}}> 0\,\,,\,\,
\ipro{u, l^{(2)}}< 0 \,\,\,.
\eeqs
In the above formulae, $l^{(1)} , l^{(2)}$ are vectors
defined in (\ref{eqn:L1},\ref{eqn:L2}) in section $3.2$.
Under the identification of the bases $\be{0} , \be{d+3}$ with divisors of
${\bp}_{\poly}$, the $K_u$ can be interpreted as the {\kae} cone in
this model.

\subsection[The correlation functions]
{Correlation functions associated with the {\kae} moduli}

\pr
D point correlation functions associated with the {\kae} class of the
{\mfd} ${\cf}({\poly})$ can be derived by using the mirror maps and
the correlation functions ${\w{a,b}}\,\,(a+b=d)$ given in section $4.1$.
These correlation functions on the {\kae} moduli space
have quantum corrections, which are expected
to be interpreted in the geometrical terms.
Firstly we want to find the variable ${\tilde{t}}_i \,\,(i=1,2)$
associated with the integral basis $h_i \in H^{1,1} ({\mm},{\bz})$.
With this basis, the {\kae} form are expanded,
\[
K_u ={\tilde{t}}_1 h_1 + {\tilde{t}}_1 h_2 \,\,\,.
\]
{}From the results in the previous subsection, the {\kae} cone $K_u$ is
also written,
\beqs
K_u =\frac{1}{2(d+1)}\ipro{u,2 {l^{(1)}}+{l^{(2)}}}\,\be{0}+
\frac{1}{2}\ipro{u, l^{(2)}}\,\be{d+3} \,\,\,.
\eeqs
Recall the asymptotic behaviour of the mirror maps in the large radius limit,
\beqs
&&{t_1}\sim\frac{1}{2 \pi i}\log x
=\frac{1}{2 \pi i} \log \frac{{b_1}{b_2}\cdots {b_{d-1}}{b_d}{b_{d+3}}}
{{b_0}^{d+1}}\,\,\,,\\
&&{t_2}\sim\frac{1}{2 \pi i}\log y
=\frac{1}{2 \pi i} \log \frac{{b_{d+1}}{b_{d+2}}}{{b_{d+3}}^2}\,\,\,.
\eeqs
We impose the asymptotic relations of the piecewise linear function $u$,
\[
u_i = \log b_i \,\,\,\,(i=0,1,2,\cdots ,{d+2},{d+3})\,\,\,.
\]
With this Ansatz , one can rewrite the {\kae} cone $K_u$,
\beqs
{K_u }&=&2 \pi i \left\{{\frac{1}{12}(2 t_1 + t_2)\,\be{0}
+\frac{1}{2} t_2 \,\be{d+3}}\right\}\\
&=& {\tilde{c}}_1 {\tilde{t}}_1 \be{0} + {\tilde{c}}_2 {\tilde{t}}_2 \be{d+3}
\eeqs
by introducing some constants ${\tilde{c}}_1 $ and ${\tilde{c}}_2 $.
In short, we get the correspondence between
the mirror maps $t_1 , t_2 $ and the variables ${\tilde{t}}_1 , {\tilde{t}}_2 $
associated with cohomology basis in
$H^{1,1} ({({\bc})}^{d+1} \verb+\+ Z_{f_{\poly}} , {\bz})$,
\[
\left(
\begin{array}{c}
t_1 \\
t_2
\end{array}
\right)
=\frac{1}{2 \pi i}
\left(
\begin{array}{cc}
6 {\tilde{c}}_1 &  - {\tilde{c}}_2  \\
0 & 2 {\tilde{c}}_2
\end{array}
\right)
\left(
\begin{array}{c}
{\tilde{t}}_1 \\
{\tilde{t}}_2
\end{array}
\right)\,\,\,.
\]
We put the conjecture that the series expansion of the correlation functions
with respect to the parameters $q_j =\exp (2 \pi i {\tilde{t}}_j)$
\,\,$(j=1,2)$
should have the integral coefficients.
Considering the asymptotic behaviours of ${\tilde{t}}_1 , {\tilde{t}}_2 $,
\beqs
2(d+1) {\tilde{c}}_1 {\tilde{t}}_1 &=& 2 \pi i (2 t_1 + t_2) \approx
2 \log \frac{1}{{(2(d+1)\psi)}^{d+1}}\,\,\,,\\
2 {\tilde{c}}_2 {\tilde{t}}_2 &=& 2 \pi i t_2 \approx
-2 \log (2 \phi)\,\,\,,
\eeqs
we may choose ${\tilde{c}}_1 =1/(d+1) \,,\, {\tilde{c}}_2 =-1$ naturally.
Then we obtain the relations,
\[
\left(
\begin{array}{c}
t_1 \\
t_2
\end{array}
\right)
=\frac{1}{2 \pi i}
\left(
\begin{array}{cc}
1 & 1  \\
0 & -2
\end{array}
\right)
\left(
\begin{array}{c}
{\tilde{t}}_1 \\
{\tilde{t}}_2
\end{array}
\right)\,\,\,,
\]
or
\[
\left(
\begin{array}{c}
{\tilde{t}}_1 \\
{\tilde{t}}_2
\end{array}
\right)
=2 \pi i
\left(
\begin{array}{cc}
1 &  1/2  \\
0 & -1/2
\end{array}
\right)
\left(
\begin{array}{c}
t_1 \\
t_2
\end{array}
\right)\,\,\,,
\]
and determine the {\kae} cone $\sigma (K)$ as
\beqs
\sigma (K)=\left\{ {{\tilde{t}}_1 h_1 + {\tilde{t}}_2 h_2 \,\,;\,\,
{\tilde{t}}_1 + {\tilde{t}}_2 >0 \,,\, {\tilde{t}}_2 <0 }\right\}
\eeqs
with
\beqs
h_1 =\frac{1}{d+1}\,\be{0}\,\,,\,\, h_2 =- \be{d+3} \,\,\,.
\eeqs
Collecting all the above results, we can define the d point
correlation functions associated with the
${\tilde{t}}_1 \,,\, {\tilde{t}}_2 $,
\beqs
{\Kappa}_{{\tilde{t}}_{i_1} {\tilde{t}}_{i_2} \cdots
 {\tilde{t}}_{i_d} } =
\frac{1}{{(\vvp{0})}^2} \sum_{{j'_1} {j'_2}\cdots {j'_d}}
\frac{\del w_{j'_1}}{\del {\tilde{t}}_{i_1}}\cdot
\frac{\del w_{j'_2}}{\del {\tilde{t}}_{i_2}}\cdot \cdots
\frac{\del w_{j'_{d-1}}}{\del {\tilde{t}}_{i_{d-1}}}\cdot
\frac{\del w_{j'_d}}{\del {\tilde{t}}_{i_d}}\cdot
{\Kappa}_{w_{j'_1} w_{j'_2} \cdots w_{j'_{d-1}} w_{j'_d} }\,\,\,,
\eeqs
where $w_i =\{{\bx\,,\,\by}\}$.
The couplings ${\Kappa}_{w_{j'_1} w_{j'_2} \cdots w_{j'_{d-1}} w_{j'_d} }$ are
symmetric completely under the permutations of any two indices
and defined as,
\beqs
&&{\Kappa}_{\bx \bx \cdots \bx \bx}:=\w{d,0}\,\,\,,\\
&&{\Kappa}_{\bx \bx \cdots \bx \by}:=\w{{d-1},1}\,\,\,,\\
&&{\Kappa}_{\bx \bx \cdots \by \by}:=\w{{d-2},2}\,\,\,,\\
&& \cdots \\
&&{\Kappa}_{\bx \bx \by \cdots \by}:=\w{2,{d-2}}\,\,\,,\\
&&{\Kappa}_{\bx \by \cdots \by \by}:=\w{1,{d-1}}\,\,\,,\\
&&{\Kappa}_{\by \by \cdots \by \by}:=\w{0,d}\,\,\,.
\eeqs
There are useful equations one can derive from the integral representation
of the mirror maps (\ref{eqn:IR1},\ref{eqn:IR2}) in section $3.3$.
\beq
&&t\equiv {\tilde{t}}_1 =
-\frac{1}{4}+\frac{1}{8\vvp{0}}\,\int_C \frac{d\nu}{{({\sin\pi\nu})}^2}
\,\frac{\gam{(d+1)\nu +1}}{{\left[{\gam{\nu +1}}\right]}^{d+1}}\
,{(2(d+1)\psi)}^{-(d+1)\nu}
\nonumber\\
&&\,\,\,\,\,\,\,\,\,\,\,\;\;\;\;\;\times \left[{u_{\nu} ({\phi})
\cos\pi\nu + u_{\nu} ({-\phi})}\right]\,\,\,,\\
&&s\equiv {\tilde{t}}_2 =
-\frac{1}{4}+\frac{1}{8\vvp{0}}\,\int_C \frac{d\nu}{{({\sin\pi\nu})}^2}
\,\frac{\gam{(d+1)\nu +1}}{{\left[{\gam{\nu +1}}\right]}^{d+1}}\,
{(2(d+1)\psi)}^{-(d+1)\nu}
\nonumber\\
&&\,\,\,\,\,\,\,\,\,\,\,\;\;\;\;\;\times \left[{u_{\nu} ({\phi})
\cos\pi\nu - u_{\nu} ({-\phi})}\right]\,\,\,.
\eeq
It follows that,
\beq
\frac{-1}{{({2(d+1)\psi})}^{d+1}}&=& q \exp \Biggl\{
-\frac{1}{\vvp{0}}
\sum^{\infty}_{n=1} \sum^{[{n/2}]}_{r=0} \frac{((d+1)n)!}{{(n!)}^d}
\,{\left\{{\frac{-1}{{({2(d+1)\psi})}^{d+1}}}\right\}}^n
\frac{{({2\phi})}^{n-2r}}{{(r!)}^2 {(n-2r)}!}\nonumber\\
&& \times \left\{{(d+1) \Psi ((d+1)n+1)-d \Psi (n+1) - \Psi (r+1)}\right\}
\Biggr\}\label{eqn:isp}\,\,\,,\\
{\phi}&=&\cosh\Biggl\{{
s+\frac{\sqrt{{\phi}^2 -1}}{2\vvp{0}} \,
\sum^{\infty}_{n=1} \frac{((d+1)n)!}{{(n!)}^{d+1}}
{\left\{ {\frac{-1}{{({2(d+1)\psi})}^{d+1}}} \right\}}^n
\,\tilde{f}_n}\Biggr\}\label{eqn:ihp}\,\,\,,\\
q&:=&{e^t} \nonumber\,\,\,,
\eeq
where the progression $\tilde{f}_n$ is defined by the recurrence relation,
\beqs
&&n \tilde{f}_n =2(2n-1) \phi \tilde{f}_{n-1} -4(n-1)({{\phi}^2 -1})
\tilde{f}_{n-2}\,\,\,(n \geq 2)\,\,,\\
&&\,\,\mbox{with}\,\,\,\, \tilde{f}_0 =0 \,,\, \tilde{f}_1 =4\,\,\,.
\eeqs
We can solve the simultaneous equation (\ref{eqn:isp},\ref{eqn:ihp})
iteratively and can express $\frac{-1}{{(2(d+1){\psi})}^{d+1}}$ and
$\phi$ as functions with respect to the variables $q$ and $s$.
For simplicity we restrict dimension $5$ case.
\begin{eqnarray*}
\lefteqn{\frac{-1}{{(12 \psi)}^6}}\\
&=&
q - 13968\,{q^2}\,\cosh (s) \\
&&+ {q^3}\,\left( -151472592 - 20113272\,\cosh (2\,s) \right) \\
&&+ {q^4}\,\left( -10207462523904\,\cosh (s) - 575186850048\,\cosh (3\,s)
      \right)  \\
&&+ {q^5}\,\bigl( -470447935319475000 -
     351288730023768000\,\cosh (2\,s) \\
&&- 11245321006807500\,\cosh (4\,s)
      \bigr)  \\
&&+ {q^6}\,\bigl( -63496742982637936184064\,\cosh (s) -
     12377026715700759881472\,\cosh (3\,s) \\
&&- 257793953453030443392\,\cosh (5\,s) \bigr) \cdots \,\,\,,\\
\lefteqn{\phi}
&=&
\cosh (s) + q\,\left( -720 + 720\,\cosh (2\,s) \right) \\
&&+ {q^2}\,\left( -6457320\,\cosh (s) + 6457320\,\cosh (3\,s) \right)\\
&&+ {q^3}\,\left( -331383545280 + 228742377600\,\cosh (2\,s) +
     102641167680\,\cosh (4\,s) \right) \\
&&+ {q^4}\,\bigl( -18870126574818240\,\cosh (s) +
     16782401580155820\,\cosh (3\,s) \\
&&+ 2087724994662420\,\cosh (5\,s) \bigr)\\
&&+ {q^5}\,\bigl( -1229739557314841424000 +
     372633864619868532000\,\cosh (2\,s) \\
&&+ 808367768482328657856\,\cosh (4\,s) + 48737924212644234144\,\cosh (6\,s)
      \bigr) \cdots \,\,\,.
\end{eqnarray*}
Gathering the above formulae, we can calculate the five point
functions ${\Kappa}_{{\tilde{t}}{\tilde{t}}{\tilde{t}}{\tilde{t}}{\tilde{t}}}$,
\begin{eqnarray*}
\lefteqn{{\Kappa}_{ttttt}}\\
&=&
1 + 113904\,q\,\cosh (s) \\
&+&{q^2}\,\bigl( 12257897184 + 5810000400\,\cosh (2\,s) \bigr) \\
&+& {q^3}\,\bigl( 2684392242065856\,\cosh (s)
+ 273249514754496\,\cosh (3\,s) \bigr) \\
&+& {q^4}\,\bigl( 257981829874371295200 + 220117176334492091136\,\cosh (2\,s)
\\
&&+ 12304491429473532432\,\cosh (4\,s) \bigr) \\
&+& {q^5}\,\bigl( 65961563983029997185594240\,\cosh (s) \\
&&+ 15336955823470891838722080\,\cosh (3\,s) \\
&&+ 538880531227754226547104\,\cosh (5\,s) \bigr) + \cdots \,\,\,,\\
\lefteqn{{\Kappa}_{tttts}}\\
&=&
73584\,q\,{\rm Sinh}(s) + 4431056400\,{q^2}\,{\rm Sinh}(2\,s) \\
&+& {q^3}\,\bigl( 740692484764992\,{\rm Sinh}(s)
+ 223867206553536\,{\rm Sinh}(3\,s) \bigr)\\
&+& {q^4}\,\bigl( 94763126006876550528\,{\rm Sinh}(2\,s) \\
&&+ 10497751479836350992\,{\rm Sinh}(4\,s) \bigr)  \\
&+& {q^5}\,\bigl( 11679174337213718521430400\,{\rm Sinh}(s) \\
&&+ 8125530240559872426667488\,{\rm Sinh}(3\,s) \\
&&+ 472005886382041853287584\,{\rm Sinh}(5\,s) \bigr) + \cdots \,\,\,,\\
\lefteqn{{\Kappa}_{tttss}}\\
&=&
-1 + 33264\,q\,\cosh (s) \\
&+& {q^2}\,\bigl( -2091181536 + 3052112400\,\cosh (2\,s) \bigr) \\
&+& {q^3}\,\bigl( -97568302413888\,\cosh (s)
+ 174484898352576\,\cosh (3\,s) \bigr)  \\
&+& {q^4}\,\bigl( -24898684426198921440 \\
&&+ 23835103083824697216\,\cosh (2\,s) \\
&&+ 8691011530199169552\,\cosh (4\,s) \bigr) \\
&+& {q^5}\,\bigl( -3002113033926710657136768\,\cosh (s) \\
&&+ 3450494265438241725942816\,\cosh (3\,s) \\
&&+ 405131241536329480028064\,\cosh (5\,s) \bigr) + \cdots \,\,\,,\\
\lefteqn{{\Kappa}_{ttsss}}\\
&=&
-2\,\coth (s) + q\,\bigl( 47880\,{\rm Csch}(s)
- 3528\,\cosh (2\,s)\,{\rm Csch}(s) \bigr)  \\
&+& {q^2}\,\bigl( 690544440\,\coth (s)
+ 836584200\,\cosh (3\,s)\,{\rm Csch}(s) \bigr)  \\
&+& {q^3}\,\bigl( 199939297711200\,{\rm Csch}(s) \\
&&- 142093529019456\,\cosh (2\,s)\,{\rm Csch}(s) \\
&&+ 62551295075808\,\cosh (4\,s)\,{\rm Csch}(s) \bigr) \\
&+& {q^4}\,\bigl( 14369096305624811616\,\coth (s) \\
&&- 5913414261982747944\,\cosh (3\,s)\,{\rm Csch}(s) \\
&&+ 3442135790280994056\,\cosh (5\,s)\,{\rm Csch}(s) \bigr)\\
&+& {q^5}\,\bigl( 2001718581208029475806528\,{\rm Csch}(s) \\
&&- 1099251328808483969708592\,\cosh (2\,s)\,{\rm Csch}(s) \\
&&+ 256067578543639354918560\,\cosh (4\,s)\,{\rm Csch}(s) \\
&&+ 169128298345308553384272\,\cosh (6\,s)\,{\rm Csch}(s) \bigr) + \cdots
\,\,\,,\\
\lefteqn{{\Kappa}_{tssss}}\\
&=&
- 3 - 4\,{{{\rm Csch}(s)}^2}
+ q\,\bigl( 56196\,\coth (s)\,{\rm Csch}(s) -
     11844\,\cosh (3\,s)\,{{{\rm Csch}(s)}^2} \bigr)  \\
&+& {q^2}\,\bigl( -1352281068\,{{{\rm Csch}(s)}^2}
+ 3051741096\,\cosh (2\,s)\,{{{\rm Csch}(s)}^2} \\
&&+ 73556100\,\cosh (4\,s)\,{{{\rm Csch}(s)}^2} \bigr)  \\
&+& {q^3}\,\bigl( 102123927100416\,\coth (s)\,{\rm Csch}(s) \\
&&+ 19002269006736\,\cosh (3\,s)\,{{{\rm Csch}(s)}^2} \\
&&+ 18930070487664\,\cosh (5\,s)\,{{{\rm Csch}(s)}^2} \bigr) \\
&+& {q^4}\,\bigl( -5814119886687093744\,{{{\rm Csch}(s)}^2} \\
&&+ 21888592432000212852\,\cosh (2\,s)\,{{{\rm Csch}(s)}^2} \\
&&- 3500145461349429192\,\cosh (4\,s)\,{{{\rm Csch}(s)}^2} \\
&&+ 1269382907731201668\,\cosh (6\,s)\,{{{\rm Csch}(s)}^2} \bigr) \\
&+& {q^5}\,\bigl( 592313888205518158535976\,\coth (s)\,{\rm Csch}(s) \\
&&+ 1052832649994450804745144\,\cosh (3\,s)\,{{{\rm Csch}(s)}^2} \\
&&- 169749335609769672076104\,\cosh (5\,s)\,{{{\rm Csch}(s)}^2} \\
&&+ 67845487961226183377256\,\cosh (7\,s)\,{{{\rm Csch}(s)}^2} \bigr) + \cdots
\,\,\,,\\
\lefteqn{{\Kappa}_{sssss}}\\
&=&
-4\,\coth (s)
-8\,\coth (s)\,{{{\rm Csch}(s)}^2} \\
&+& q\,\bigl( -73080\,{\rm Csch}(s) - 43848\,\cosh (2\,s)\,{\rm Csch}(s)
\bigr)\\
&+& {q^2}\,\bigl( 8605688184\,\coth (s)
- 542359800\,\cosh (3\,s)\,{\rm Csch}(s) \bigr)  \\
&+& {q^3}\,\bigl( 37219523344224\,{\rm Csch}(s) \\
&&+ 353500849433280\,\cosh (2\,s)\,{\rm Csch}(s) \\
&&+ 13168986874848\,\cosh (4\,s)\,{\rm Csch}(s) \bigr)\\
&+& {q^4}\,\bigl( 29797773008663870208\,\coth (s) \\
&&+ 5789880986681623032\,\cosh (3\,s)\,{\rm Csch}(s) \\
&&+ 1635395840643812616\,\cosh (5\,s)\,{\rm Csch}(s) \bigr) \\
&+& {q^5}\,\bigl( 847125331875900500330304\,{\rm Csch}(s) \\
&&+ 3137482755359009546571216\,\cosh (2\,s)\,{\rm Csch}(s) \\
&&- 87672381735274833962592\,\cosh (4\,s)\,{\rm Csch}(s) \\
&&+ 102253653499596180124752\,\cosh (6\,s)\,{\rm Csch}(s) \bigr) + \cdots
\,\,\,,
\end{eqnarray*}
where $t \equiv {\tilde{t}}_1 $ and $s \equiv {\tilde{t}}_2 $.
The classical parts of these five point correlation functions can be
interpreted as the intersections of some divisors. Firstly we consider
a divisor $L$, which is defined by a surface of degree $2(d+1)$ in
${\bf P}_{d+1} [2,2,2,\cdots ,2,2,1](2(d+1))$ with one parameter $\lambda$,
\[
{z_1}^{d+1} +{z_2}^{d+1} +\cdots +{z_{d-1}}^{d+1} +{z_d}^{d+1}
+ (1+ \lambda ){z_{d+1}}^{2(d+1)} =0\,\,\,.
\]
Because any two distinct elements of the linear system $|L|$ are disjoint,
we obtain $L \cdot L=0$.
Secondly we take a linear system $|H|$, which is generated by degree $2$
polynomials.
This linear system is written as
\[
|H|=|2L+E|\,\,\,,
\]
where $E$ is the exceptional divisor.
Because the variables ${z_1},{z_2},\cdots ,{z_{d-1}},{z_d}$ are degree $2$,
we express these d variables as some quadratic homogeneous
polynomials in $z_{d+1}$ and $z_{d+2}$ in the calculation of an intersection
$H^d$. In such an calculation, we have a homogeneous polynomial
of degree $2(d+1)$ in the variables $z_{d+1}$ and $z_{d+2}$ and we get
$H^d =2(d+1)$. By using the similar argument, one can read
$ H^{d-1} \cdot L=d+1$. Using the relations,
\beqs
&&H^d =2(d+1)\,\,,\,\, H^{d-1} \cdot L=d+1\,\,,\,\, L^2 =0\,\,,\\
&&H=2L+E\,\,,
\eeqs
we can calculate intersection numbers of the divisors $H$ and $L$,
\beqs
&&H^d =2(d+1)\,\,,\,\,H^{d-1} \cdot E=0\,\,,\,\,
H^{d-2} \cdot E^2 =-2(d+1)\,\,,\cdots \\
&&H^2 \cdot E^{d-2} =-2(d+1)(d-3)\,\,,\,\,H \cdot E^{d-1}
=-2(d+1)(d-2)\,\,,\,\,E^d =-2(d+1)(d-1)\,\,.
\eeqs
Note the following equations,
\beqs
&&-2 \coth (s)=-2-4\cdot\frac{p^{-2}}{1- p^{-2}}\,\,\,,\\
&&-3-4 {\rm Csch}(s)^2 =-3-16\cdot\frac{p^{-2}}{{(1- p^{-2})}^2}\,\,\,,\\
&&-4 \coth (s)-8\coth (s){\rm Csch}(s)^2
=-4-\frac{40 p^{-2}}{{(1- p^{-2})}^2} -\frac{56 p^{-4}}{{(1- p^{-2})}^3}
-\frac{8 p^{-6}}{{(1- p^{-2})}^3}\,\,\,,\\
&&p:={e^s}\,\,\,.
\eeqs
we obtain the classical parts of the five point correlation functions in the 5
dimensinal case,
\beqs
&&{\Kappa}^0_{ttttt} =1\,\,,\,\,
{\Kappa}^0_{tttts} =0\,\,,\,\,
{\Kappa}^0_{tttss} =-1\,\,,\,\,\\
&&{\Kappa}^0_{ttsss} =-2\,\,,\,\,
{\Kappa}^0_{tssss} =-3\,\,,\,\,
{\Kappa}^0_{sssss} =-4\,\,.
\eeqs
These numbers coincide with the intersection numbers
up to a common overall factor $12$.
As for the corrections in the couplings, these are not understood in the
geometrical terms yet.

\newpage
\section{The monodromy}

\pr
In this section, we study monodromy transformations and give monodromy
matrices.

\subsection[The monodromy]{The monodromy}

\pr
The defining equation of two parameter family of the Calabi-Yau $d$-fold
was represented in (\ref{eqn:ZERO2}). If the conditions
${\del}p/{\del}{z_i}=0,\hspace{2mm} (i=1,2,\cdots ,{d+2})$ are
satisfied for some $\psi,\phi$, this variety gets singular.
These conditions are rewritten in the following four cases,\\
1.\,\,Along the locus $\phi + 2^d {\psi}^{d+1} =\pm 1$, the d-fold
$\left\{{p=0}\right\}/G$ has a collection of conifold points.\,\,$(C_{con})$\\
2.\,\,Along the locus $\phi =\pm 1$,
the d-fold $\left\{{p=0}\right\}/G$ has
isolated singularities.\,\,$(C_1)$\\
3.\,\,Along the locus $\psi =0$, the d-fold
$\left\{{p=0}\right\}/G$ gets singular.
\,\,$(C_0)$ \\
4.\,\,When the parameters $\phi$ and $\psi$ tend to infinity,
one gets a singular fold $(C_{\infty})$,
\[
{\left\{{2(d+1)\psi {z_1}{z_2}\cdots {z_d}{z_{d+1}}{z_{d+2}}+2 \phi
{z^{d+1}_{d+1}}{z^{d+1}_{d+2}}=0
}\right\}/G}\,\,\,.
\]
\,\,
The group $G$ is characterized by its elements $g \in G$,
\[
g=({{\alpha}^{2 a_1} ,{\alpha}^{2 a_2} ,\cdots ,{\alpha}^{2 a_{d-1}} ,
{\alpha}^{2 a_d} ,{\alpha}^{a_{d+1}} ,{\alpha}^{a_{d+2}}})\,\,\,,\,\,
\alpha := \exp \frac{2 \pi i}{2(d+1)}\,\,\,,
\]
acting on $({{z_1},{z_2},\cdots ,{z_{d-1}},{z_d},{z_{d+1}},{z_{d+2}},
\psi ,\phi})$
as
\[
({{\alpha}^{2 a_1} {z_1} ,{\alpha}^{2 a_2} {z_2} ,\cdots
,{\alpha}^{2 a_4} {z_{d-1}} ,{\alpha}^{2 a_d} {z_d} ,{\alpha}^{a_{d+1}}
{z_{d+1}}
,{\alpha}^{a_{d+2}} {z_{d+2}},{\alpha}^{-a} {\psi},{\alpha}^{-(d+1)a}
{\phi}})\,\,\,,
\]
with $a:=2 {a_1}+2 {a_2}+\cdots +2 {a_{d-1}}+2 {a_d}+{a_{d+1}}+{a_{d+2}}$.
This group $G$ acts on the parameter space $\left\{{({\psi},{\phi})}\right\}$
as
\beq
({\psi},{\phi}) \rightarrow ({\alpha}{\psi},-{\phi})\label{eqn:trans}\,\,\,.
\eeq
If one rotates the values of the parameters $({\psi},{\phi})$ around the
special values defined in 1.,2.,3.,4. analytically,
homology cycles ${\gamma}_i$ of the variety $\left\{{p=0}\right\}/G$
turn into linear combinations of other homology cycles
${\displaystyle \sum_j c_{ij} {\gamma}_j}$. As a consequence,
the periods also change into linear combinations of others.
The matrices associated with this linear transformations are called
monodromy matrices and contain geometrical information.
Recall the discriminant loci ${\poly}_1 ,{\poly}_2 ,{\poly}_3 $ and
${\poly}_4 $ in section $4.1$. These can be rewritten as
\beqs
&&{\poly}_1 = {(1-\bx)}^2 - {\bx}^2 \,\by =
\frac{1}{{(2^d {\psi}^{d+1})}^2} ({2^d {\psi}^{d+1} +\phi +1})\,
({2^d {\psi}^{d+1} +\phi -1})\,\,\,,\\
&&{\poly}_2 =1-\by =\frac{1}{{\phi}^2}\,
({\phi +1})({\phi -1})\,\,\,,\\
&&{\poly}_3 =\bx = -\frac{\phi}{2^d {\psi}^{d+1}}\,\,\,,\\
&&{\poly}_4 =\by =\frac{1}{{\phi}^2}\,\,\,.
\eeqs
The zero loci of ${\poly}_1$ are $\phi + 2^d {\psi}^{d+1} =\pm 1$
and associate with the case $1$.
Secondly the zero loci of ${\poly}_2$ are $\phi =\pm 1$
and correspond to the case $2$.
Thirdly the discriminant ${\poly}_3 =\bx $ is the fixed point of the
transformation (\ref{eqn:trans}).
Lastly the discriminant ${\poly}_4$ have zero at $\phi =\infty$ and
corresponds to the case $4$.

\subsection[The monodromy]{Monodromy matrices}

\pr
In this subsection, we treat the monodromy matrices of the fivefold.
In order to obtain the monodromy matrices, let us take a basis $v$ of the
periods,
\[
v:= {}^t ({\vvp{0}},{\vvp{1}},{\vvp{2}},{\vvp{3}},{\vvp{4}},{\vvp{5}},
{\vvp{6}},{\vvp{7}},{\vvp{8}},{\vvp{9}})\,\,\,.
\]
Firstly we consider the conifold singularity along the locus
$\phi +2^5 {\psi}^6 =1$.
By means of the same argument made in {\cite{COGP}},
it can be understood that the periods have the structure,
\[
\vvp{j} ({\psi},{\phi})= \frac{c_j}{2 \pi i}
({{\vvp{1}}-{\vvp{0}}}) \,\log ({2^5 {\psi}^6 -\phi -1}) +f_j
({\psi},{\phi})\,\,\,,
\]
where the $f_j$ is an analytic functions in the neighbour of this singular
point and the $c_j$ are some constant coefficients.
The monodromy transformation ${\cal T}$ of this
singularity acts on the periods,
\beqs
{\cal T}\vvp{j} =\vvp{j} +{c_j}({{\vvp{1}}-{\vvp{0}}})\,\,\,.
\eeqs
We obtain the coefficients $c_j$,
\beqs
&&{c_0}=-1\,,\,{c_1}=-1\,,\,{c_2}=1\,,\,{c_3}=5\,,\,{c_4}=-5\,,\,\\
&&{c_5}=-10\,,\,{c_6}=10\,,\,{c_7}=10\,,\,{c_8}=-10\,,\,{c_9}=-5\,\,.
\eeqs
The monodromy matrix $T$ associated with ${\cal T}$ are found,
\[
T=\left(
\begin{array}{rrrrrrrrrr}
  2 &  -1 &  0  &  0  &  0  &  0  &  0  &  0  &  0  &  0   \\
  1 &   0 &  0  &  0  &  0  &  0  &  0  &  0  &  0  &  0   \\
 -1 &   1 &  1  &  0  &  0  &  0  &  0  &  0  &  0  &  0   \\
 -5 &   5 &  0  &  1  &  0  &  0  &  0  &  0  &  0  &  0   \\
  5 &  -5 &  0  &  0  &  1  &  0  &  0  &  0  &  0  &  0   \\
 10 & -10 &  0  &  0  &  0  &  1  &  0  &  0  &  0  &  0   \\
-10 &  10 &  0  &  0  &  0  &  0  &  1  &  0  &  0  &  0   \\
-10 &  10 &  0  &  0  &  0  &  0  &  0  &  1  &  0  &  0   \\
 10 & -10 &  0  &  0  &  0  &  0  &  0  &  0  &  1  &  0   \\
  5 &  -5 &  0  &  0  &  0  &  0  &  0  &  0  &  0  &  1
\end{array}
\right)\,\,\,.
\]
Secondly let us consider the isolated singularity along the locus $\phi =1$.
By the monodromy transformation ${\cal B}$ of this singularity,
the function $u_{\nu} ({\phi}),u_{\nu} ({-\phi})$
defined in (\ref{eqn:UN}) are transformed linearly,
\beqs
&&{u_{\nu}}({\phi}) \rightarrow (1- e^{2\pi i \nu})
{u_{\nu}}({\phi})+ e^{\pi i \nu} {u_{\nu}}({-\phi})\,\,\,,\\
&&{u_{\nu}}({-\phi}) \rightarrow e^{\pi i \nu} {u_{\nu}}({\phi})\,\,\,.
\eeqs
Using the integral representation of the periods with the change of
the contour $C$ to enclose the poles $\nu =0,1,2,\cdots$ clockwise,
we obtain the monodromy matrix $B$ associated with ${\cal B}$,
\[
B=\left(
\begin{array}{rrrrrrrrrr}
  1 &   0 &   0  &   0  &   0  &   0  &   0  &   0  &   0  &   0   \\
  0 &   1 &  -1  &   1  &  -1  &   1  &  -1  &   1  &  -1  &   1   \\
  0 &   0 &   2  &  -1  &   1  &  -1  &   1  &  -1  &   1  &  -1   \\
  5 &  -5 &   6  &  -5  &   5  &  -5  &   5  &  -5  &   5  &  -5   \\
 -5 &   5 &  -6  &   6  &  -4  &   5  &  -5  &   5  &  -5  &   5   \\
-10 &  10 & -10  &  10  &  -9  &  10  & -10  &  10  & -10  &  10   \\
 10 & -10 &  10  & -10  &   9  &  -9  &  11  & -10  &  10  & -10   \\
 10 & -10 &  10  & -10  &  10  & -10  &  11  & -10  &  10  & -10   \\
-10 &  10 & -10  &  10  & -10  &  10  & -11  &  11  &  -9  &  10   \\
 -5 &   5 &  -5  &   5  &  -5  &   5  &  -5  &   5  &  -4  &   5
\end{array}
\right)\,\,\,.
\]
Thirdly the monodromy matrix $A$ associated with the monodromy transformation
${\cal A}$ along the locus $\psi =0$, \,
$i.e.\,({\psi},{\phi})\rightarrow ({\alpha}{\psi},{-\phi})$
can be obtained easily,
\[
A=\left(
\begin{array}{cccccccccc}
  0 &   1 &  0  &  0  &  0  &  0  &  0  &  0  &  0  &  0   \\
  0 &   0 &  1  &  0  &  0  &  0  &  0  &  0  &  0  &  0   \\
  0 &   0 &  0  &  1  &  0  &  0  &  0  &  0  &  0  &  0   \\
  0 &   0 &  0  &  0  &  1  &  0  &  0  &  0  &  0  &  0   \\
  0 &   0 &  0  &  0  &  0  &  1  &  0  &  0  &  0  &  0   \\
  0 &   0 &  0  &  0  &  0  &  0  &  1  &  0  &  0  &  0   \\
  0 &   0 &  0  &  0  &  0  &  0  &  0  &  1  &  0  &  0   \\
  0 &   0 &  0  &  0  &  0  &  0  &  0  &  0  &  1  &  0   \\
  0 &   0 &  0  &  0  &  0  &  0  &  0  &  0  &  0  &  1   \\
 -1 &   0 & -1  &  0  & -1  &  0  & -1  &  0  & -1  &  0
\end{array}
\right)\,\,\,.
\]
Lastly the monodromy transformation at the infinity is expressed
${({\cal BTA})}^{-1}$.
Thus the monodromy matrix can be read,
\[
{(ATB)}^{-1} =\left(
\begin{array}{cccccccccc}
  1 &   0 &  0  &  0  &  0  &  0  &  0  &  0  &  0  &  0   \\
  1 &   1 & -1  &  0  &  0  &  0  &  0  &  0  &  0  &  0   \\
  0 &   0 &  1  &  0  &  0  &  0  &  0  &  0  &  0  &  0   \\
  0 &   0 &  1  &  1  & -1  &  0  &  0  &  0  &  0  &  0   \\
  0 &   0 &  0  &  0  &  1  &  0  &  0  &  0  &  0  &  0   \\
  0 &   0 &  0  &  0  &  1  &  1  & -1  &  0  &  0  &  0   \\
  0 &   0 &  0  &  0  &  0  &  0  &  1  &  0  &  0  &  0   \\
  0 &   0 &  0  &  0  &  0  &  0  &  1  &  1  & -1  &  0   \\
  0 &   0 &  0  &  0  &  0  &  0  &  0  &  0  &  1  &  0   \\
  1 &   0 &  1  &  0  &  1  &  0  &  1  &  0  &  2  &  1
\end{array}
\right)\,\,\,.
\]

\newpage


\section{Conclusion}
\pr
In this paper, we constructed a two parameter family of the Calabi-Yau
d-fold by the method of the toric variety. The correlation functions
associated with the complex moduli in the mirror manifold are
meromorphic functions and become singular when at least one of the
discriminant loci ${\poly}_1 ,{\poly}_2 ,{\poly}_3 ,{\poly}_4 $ vanishes.
These discriminant loci correspond to the singular points of the manifold in
secton $5$ and these singularities are reflected in the correlation
functions associated with the complex moduli space.
By contrast, the correlation functions associated with the {\kae} moduli
in the original manifold have quantum corrections. The classical parts
of these {\kae} couplings can be interpreted in geometrical terms.
Quantum corrections in these couplings in d-fold have more
complicated form than threefolds. In the case of Calabi-Yau threefolds,
there are isolated rational curves in these threefolds and quantum
corrections in the three point couplings associated with the {\kae}
moduli can be interpreted as the numbers of these curves.
As already indicated for the one parameter models {\cite{GMR}},
there are families of rational curves in higher $(\geq 4)$ dimensions.
So our results seem to reflect families of instantons with continuous
parameters more manifestly than cases of the one parameter models
and it is expected to be interpreted as intersection numbers of
some Chern class with {\kae} forms.


\end{document}